# West-east asymmetry of a mini-magnetosphere induced by Hall effects

**I F Shaikhislamov, V G Posukh, A V Melekhov, Yu P Zakharov, E L Boyarintsev and A G Ponomarenko**

Dep. of Laser Plasma, Institute of Laser Physics SB RAS, pr. Lavrentyeva 13/3, Novosibirsk, 630090, Russia,

e-mail: ildars@ngs.ru

*Abstract:* Magnetosphere comparable in size to ion inertia length, or a mini-magnetosphere, possesses unusual features as was predicted by numerical simulations and shown in recent experiments. In the present paper we study a pronounced difference between west and east flanks of mini-magnetosphere observed for the first time in pioneering Terrella experiments of the 60-s. It manifests itself in plasma penetration deep inside of the west flank and formation of a return current that effect magnetic structure, in contrast to the east flank with well defined boundary layer and plasma cavity. We propose that the plasma penetration and the return current can be understood in the frame of a test particle model. This model serves to illustrate in simple terms the Hall physics which was found to be behind similar features at the frontal part and the tail of mini-magnetosphere observed in our previous experiments. Because of large gyroradius ions tend to be deflected by dipole field at the east flank and drawn into the west flank. To verify that the return electric current in magnetospheric plasma is carried by ions we measured it directly by Rogovski coil and compared with ion flux measured by Langmuir probe.

## 1. Introduction

Mini-magnetosphere defined as a magnetosphere with size of the order of ion inertia length is a novel subject in space plasma research. It might form around small bodies like an asteroid with a remnant magnetic field, or above magnetic anomalies like on Moon or Mars. A subject of mini-magnetosphere is also of interest to future applications of magnetic field sources on-board a spacecraft for the purpose of crew protection from energetic galactic protons or propulsion by means of magnetoplasma sail. The studies were initiated by Galileo spacecraft encounter with the asteroid Gaspra in 1991 and Ida in 1993, and it was immediately realized that Hall physics is crucial for understanding plasma effects around such objects (*Kivelson et al 1993*). The first involved treatment of a mini-magnetosphere was based on numerical studies by Hall MHD and hybrid codes. It was found out that at the front there is no ion deflection and density pile up. A shocked upstream region and a strong obstacle to Solar Wind roughly resembling magnetospheric bowshock appear only when pressure balance stand off distance is larger than the ion inertia scale (*Omidi et al 2002*). Hybrid simulations showed that the size of a mini-magnetosphere is equal to MHD stand off distance when ion inertia length is small and to a closest approach distance of test particles in dipole field when it is large (*Fujita, 2004*). It should be noted, however, that findings of numerical simulations remain somewhat controversial as in the number of works, for example (*Harnett & Winglee 2003, Kajimura et al 2006, Gargate et al 2008*) no significant departure from the usual MHD was reported.

     In our previous papers (*Shaikhislamov et al 2013, 2014*) a mini-magnetosphere was studied by means of laboratory experiments, 2.5D Hall MHD numerical simulation and theoretical analysis. In

particular, experiments verified in details that when ion inertia length is larger than the pressure balance distance the plasma penetrates into magnetosphere and is stopped at the Stoermer limit of minimum approach of test ions. The other new feature not reported in simulations was that in dependence on ion inertia length magnetopause shifts farther away from the pressure balance distance and jump of field at magnetopause lessens. Based on experimental data a comprehensive model was built for the first time which explains most important features of mini-magnetosphere observed in (*Shaikhislamov et al 2013*) and in other simulations sited above. Non-coplanar component of magnetic field directed perpendicular to dipole components was found to be related to such dramatic change. Experimentally observed spatial structure and independence on the sign of dipole magnetic moment gave direct evidence that this field is generated by the Hall term. Quantitative analytical estimates of sub-solar magnetopause position, plasma penetration velocity and Hall field in dependence on ion inertia length were shown to be consistent with results of numerical simulation and experimental data. Developed model explains why a mini-magnetosphere is so much different. At magnetopause the Chapman-Ferraro current generates non-coplanar magnetic field along its direction as described by the Hall term $\mathbf{J} \times \mathbf{B}/nec$ in the Ohm' law. In the noon-midnight meridian plane of GSM frame this is out of plane component $\mathbf{B}_y$. The resulting new system of Hall current $J_x \approx -(c/4\pi)\partial B_y/\partial z$ advects magnetic field, as described by the same Hall term. In steady state to cancel such additional advection the plasma velocity tends to balance the current velocity. In other words, Hall current and advection combined tend to cancel induction electric field $\mathbf{E} \approx \mathbf{J} \times \mathbf{B}/n_e ec - \mathbf{V}_i \times \mathbf{B}/c \approx 0$. This in effect means three things. First, plasma penetrates into magnetosphere, and because the jump of kinetic pressure lessens the magnetopause position correspondingly shifts away from dipole. Such crucial fact as disappearance of bowshock is also explained by penetration of plasma across magnetopause. With increase of Hall current penetration velocity also increases and, when it exceeds maximum possible velocity in magnetosheath region as determined by Rankine-Hugoniot relations, a standing shock cannot exist. Second, plasma dynamics inside of magnetosphere is described by a particle motion law in the dipole field, and in such a case plasma is stopped at the Stoermer limit of ions. Third, the Hall current inside of magnetosphere is carried by ions $\mathbf{J} \approx ez_i n_i \mathbf{V}_i$.

Numerical simulation in (*Shaikhislamov et al 2013*) also revealed some other counter-intuitive features of mini-magnetosphere. First, upstream electrons bypass magnetosphere around magnetopause boundary and don't directly penetrate inside as ions do, while inside of magnetosphere a motionless population of electrons exists which neutralize ion flow. Indeed, as the magnetic field is frozen into electron fluid, in a stationary case electrons should be at rest inside of magnetosphere, which is equivalent to induction electric field being zero. Thus, if indirect processes of exchange are slow enough, magnetospheric population of electrons might develop features distinctly different from SW electrons and this can be of fundamental and practical interest. Second, without Interplanetary Magnetic Field the lobe magnetic field sufficiently far in the tail is dominated by out of plane Hall component rather than by the tail-ward directed field. This is of practical interest for a spacecraft crossing and data interpretation as well. In the follow up experiment (*Shaikhislamov et al 2014*) it was indeed found out that in the tail lobes the perpendicular Hall field dominates over dipolar components. Additionally, distinct low-hybrid oscillations have been observed in regions where, according to the model, a population of electrons at a rest should exists in order to neutralize ions moving across magnetosphere.

Despite the progress based on laboratory experiments a whole 3-dimension picture of mini-magnetosphere remains elusive. The reason is that measurements in those experiments were made so far in the meridian plane of magnetosphere. It should be noted that our simulations as well as pioneering numerical study by (*Omidi et al 2002*) have been made in 2.5D geometry corresponding to meridian plane. The quantitative estimates of the model developed in (*Shaikhislamov et al 2013*) also are restricted by the meridian plane of magnetosphere. Correspondingly, the first dedicated experiments have been tailored to the same geometry.

Interestingly, the structure of mini-magnetosphere in a terminator plane (Y-Z in GSM frame), or the structure of the west and east flanks, have a long history of research and started in pioneering Terrella experiments of the 60-s. In (*Cladis et al. 1964*, *Kawashima and Mori 1965*) a significant west-east asymmetry was revealed. Inside of the west flank of magnetosphere a strong current was measured which flows reverse to the Chapman-Ferraro current. This resulted in larger field variation at magnetopause and field decrease inside of magnetosphere. Moreover, plasma freely penetrated inside of the west flank, while in the east flank no such unusual features were observed. The same features are reported in later experiments of (*Ponomarenko et al 2004*) where detailed spatial distribution of plasma and magnetic field have been obtained showing a well defined magnetopause and plasma cavity at the front of the dipole and the east flank, and contrary to these a distinct region of large plasma density inside of the west flank accompanied by decrease of magnetic variation. Review of Terrella experiments from the point of view of a mini-magnetosphere can be found in (*Antonov et al 2014*).

The aim of our latest experiment reported in the present paper was to investigate the west-east asymmetry of mini-magnetosphere and to obtain evidence that it has the same Hall nature as the features observed in the meridian plane, such as ions penetration and accompanying Hall current. Applying this hypothesis to the 3-dimentional magnetosphere, one can easily perceive that due to gyrorotation in dipole field ions tend to penetrate inside of the west flank and to deflect from the east flank of magnetosphere. At the same time, ions directly flowing past dipole should generate inside of the west flank a current in direction matching the return current, which is the main course of asymmetry of mini-magnetosphere reported in (*Cladis et al. 1964, Kawashima and Mori 1965, Ponomarenko et al 2004*). Thus, if as well the value of return current will turn to be equal to ion current $ez_i n_i \mathbf{V}_i$ then it will be a direct confirmation of the proposed hypothesis.

However, dissimilar to meridian plane, in the case of Hall current being parallel to Chapman-Ferraro current it is difficult to differentiate between them as they both produce magnetic field coplanar to dipole components. Inversion of dipole moment also doesn't discriminate between those currents as they both don't change sign in the terminator plane. This is because moment inversion not only reverses the dipole magnetic field but interchanges west and east flanks as well. Calculation of one component of current density from magnetic field measurements requires detailed 2D mapping, and the result is an average of many shots which lessens an accuracy.

To make direct, and thus more reliable, measurement of electric current density we employed Rogovski coil developed specifically to satisfy rigorous conditions of plasma environment. Its performance was checked in measurements at the frontal part of mini-magnetosphere well known from previous experiments. The magnetic and plasma measurements at the flanks of mini-magnetosphere fully confirmed the results of previous studies on the effects of asymmetry (*Cladis et al. 1964, Kawashima and Mori 1965, Ponomarenko et al 2004*). Direct measurements of electric current and ion flux confirmed that the return current inside of the west flank is close in value to ion current. This fact was also confirmed by estimation of electric current density from magnetic measurements. Thus, we deduce that mini-magnetosphere asymmetry has the same Hall nature as the plasma penetration across sub-solar magnetopause.

The paper consists of four sections. In the second section a structure of mini-magnetosphere predicted by test particle model is briefly outlined. The third section contains experimental set up and results followed by conclusions.

## 2. Test particle model

To illustrate the idea of the paper and to prepare a context for experimental results presented below we briefly outline the structure of inner magnetosphere that can be derived in frame of the test particle model. Keeping the problem in the equatorial plane (X-Y) of magnetic dipole makes it very simple for numerical treatment. Further on we use the GSM coordinate system with dipole placed at the origin. We apply the Stoermer approach to mini-magnetosphere by injecting particles from imaginary boundary assuming that dipole field is spatially restricted by the magnetopause. Outside this fixed boundary the particles of upstream flow move in straight lines along X axis. The only dimensionless parameter of the problem – relation of Stoermer radius to the sub-solar distance of magnetosphere boundary – was taken according to conditions of experiment. Example of trajectories is shown in fig. 1A. One can see how significant the west and east flanks are different when Stoermer radius is comparable to magnetosphere size. In the east flank ions are deflected after entering the region of dipole field, while due to the same Larmor rotation they penetrate deep into the west flank of magnetosphere. Using sufficiently large number of test ions a macroscopic distribution of density and fluid velocity can be calculated. Profile along Y axis (fig. 1B) demonstrates that at the east flank (Y>0) a clear cavity devoid of plasma with size comparable to frontal magnetopause distance exist. However, except small region near the dipole, the west flank (Y<0) is filled by plasma close in density and velocity to undisturbed upstream flow. The qualitative validity of the test particle model has been demonstrated in our earlier paper (*Shaikhislamov et al 2013*) and is based on the fact that due to Hall current electric field in the inner part of mini-magnetosphere is negligible and ions move only under magnetic force. The quantitative applicability of the model is rather reduced because of finite width of boundary layer where electric field isn't zero and because the form of magnetopause boundary depends on the Hall effects as well.

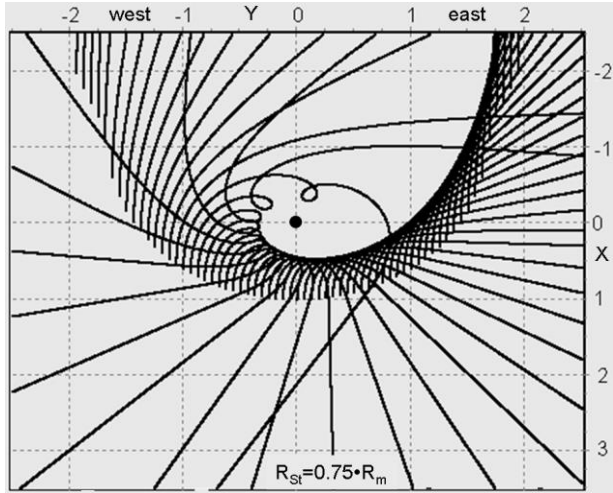 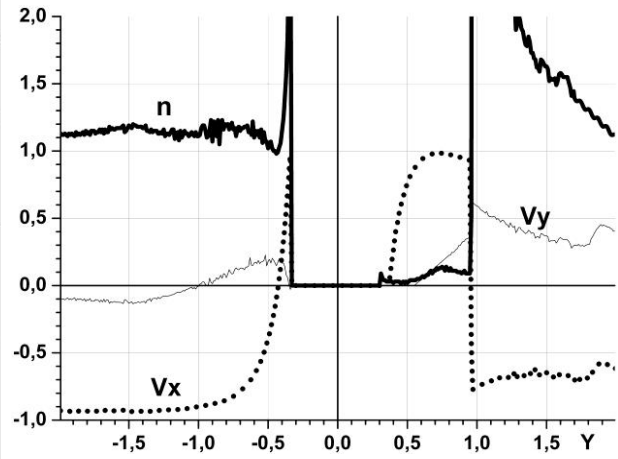

Figure 1A. Ion trajectories in equatorial plane of dipole magnetic field which is restricted by imaginary magnetopause.

Figure 1B. West-east profiles of normalized ion density and velocity calculated by test particle model.

## 3. Experimental set up and results

Set up and conditions are practically the same as in our previous experiments (*Shaikhislamov et al 2013, 2014*). Plasma is generated by theta-pinch. Discharged hydrogen ions flow along axis of vacuum chamber 5 m in length and 1.2 m in diameter. At the chamber center magnetic dipole oriented transverse to the flow is placed. Dipole epoxy cover has a diameter of 5 cm. Operating time of theta-pinch and dipole is $50\,\mu s$ and $200\,\mu s$ respectively. After a time of about $20\,\mu s$ following discharge a quasi-steady state laboratory magnetosphere with spatial scale of the order of 10 cm is formed. Geometry of experiment is shown in fig. 2 while specific conditions and parameters are listed in the Table 1. Large value of plasma kinetic scales relative to the size of magnetosphere was achieved by lowering plasma density and magnetic moment. For the given ion inertia length $L_{pi} = c/\omega_{pi}$ and the calculated pressure balance stand off distance $R_M = \left(\mu^2/2\pi n_i M V_o^2\right)^{1/6}$ the realized Hall parameter was distinctly smaller than unity $D = R_M/L_{pi} \approx 0.8$. Like the SW, the flow is super-sonic ($M_s \approx 3$) and sufficiently collisionless. Note that there was no frozen-in magnetic field as analog of IMF.

Diagnostics consisted of Langmuir electric probe and three orthogonal magnetic coils contained in a single probe head about 1 cm in length and 0.5 cm in diameter. Two such heads were fasten to a two-pronged device. Most measurements have been made in equatorial plane either in front of the dipole or at the flanks in the terminator cross-section. Reversing the dipole moment in effect switches the west and east flanks of magnetosphere at which the probes are positioned. To measure current density in plasma Rogovski coil was employed. It consisted of two parallel 1000-turns coils put inside of dielectric tube 5 mm in diameter. Coils have been insulated from outside electric fields by copper foil with slit going parallel to the container tube. The tube was bent to form a full circle with diameter of 6 cm. Coils have been connected in reverse to measure opposite magnetic fluxes. The measurement was obtained by mathematical deduction of signals from opposite but otherwise identical coils to further reduce obstructive interference of quasi-static potential or high frequency turbulent electric fields in plasma.

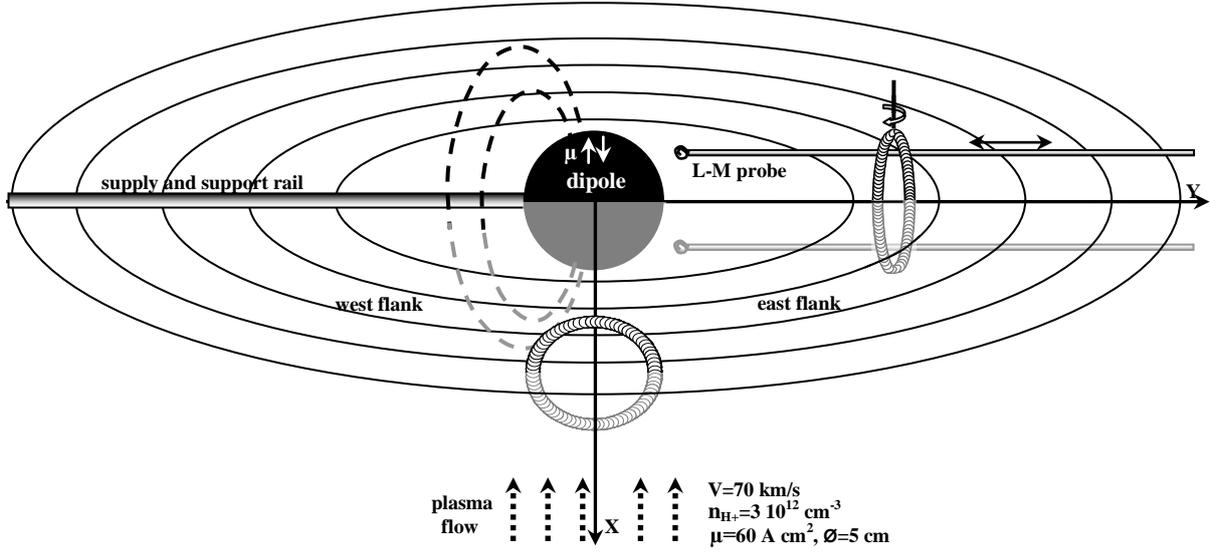

Figure 2. Experimental set up.

Table 1. Parameters of experiment.

| parameter | value | parameter | value |
|---|---|---|---|
| density $n_{H+}$ | $3 \cdot 10^{12}$ cm$^{-3}$ | stand off $R_M$ | 10 cm |
| velocity $V_o$ | 70-100 km/s | Hall D | 0.8 |
| elec. temp. $T_e$ | 5-10 eV | Sonic Mach | ~3 |
| mag. moment $\mu$ | 60 A·m$^2$ | Knudsen Num. | >100 |
| dipole rad. $R_d$ | 2.5 cm | Reynolds Num. | ~100 |

Fig. 3 demonstrates typical probe signals measured close to the dipole. Zero time approximately corresponds to arrival of plasma flow to the dipole location. The ion current collected by Langmuir probe is given by $J_p = en_e V_i \sqrt{1 + 2|U_c|/m_i V_i^2}$ where $V_i$, $m_i$ are velocity and mass of ions. For the theta-pinch hydrogen plasma with typical proton energy being much smaller than used ion collector potential $U_c = -95$ V the probe operates in a regime of electron density measurement $n_e(\text{cm}^{-3}) \approx 4.5 \cdot 10^{11} J_p(\text{A}/\text{cm}^2)$ which is shown in fig. 3. Velocity of the flow was measured by means of two widely separated Faraday cups and equaled to about of $V_{io} = 70$ km/s. Using this velocity calculated ion current $J_{ion} = en_e V_i$ is shown in fig. 3 as well. Quasi-stationary magnetosphere is sustained from about 15 μs to 40 μs, as indicated by vertical lines, which is about 20 characteristic magnetospheric times $R_M/V_o$. During this interval there are some oscillations of plasma flow and magnetospheric compression generated by theta-pinch discharge. Fig. 3 also presents an example of Rogovski coil measurement close to the sub-solar magnetopause location at a distance of X=11 cm when the dipole was switched on. It was oriented with normal vector along Y axis so as to measure Chapman-Ferraro current which forms magnetosphere. The shown result is obtained by time integrating of actual signal and dividing it by Rogovski coil cross-section.

The frontal structure of mini-magnetosphere obtained by measurements along X axis is shown in fig. 4. Further on, all signals used for spatial plots were averaged over time interval indicated in fig. 2. Magnetic variation is approximately constant inside of magnetosphere at a level slightly below 20 G. In the boundary layer it rapidly changes and crosses zero. Away from the "sub-solar" point its value corresponds to dipole field but of opposite sign meaning that outside of magnetosphere the total field $B = \delta B + B_d$ is close to zero.

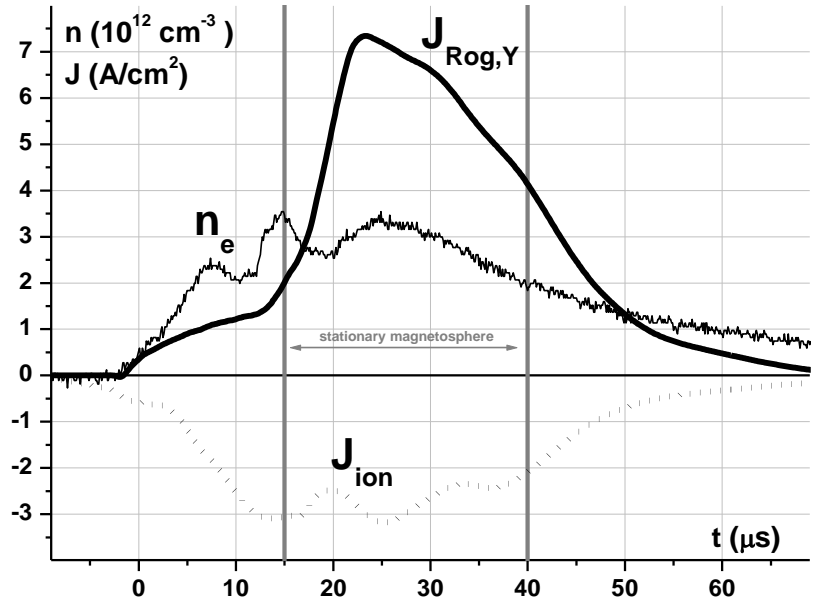

Figure 3. Time behavior of plasma density (thin solid line) and ion current (dotted line) measured by Langmuir probe at the dipole location. Straight lines indicate a time interval of quasi-stationary flow. Thick solid line – electric current density measured by Rogovski coil at a distance of X=11 cm with dipole switched on.

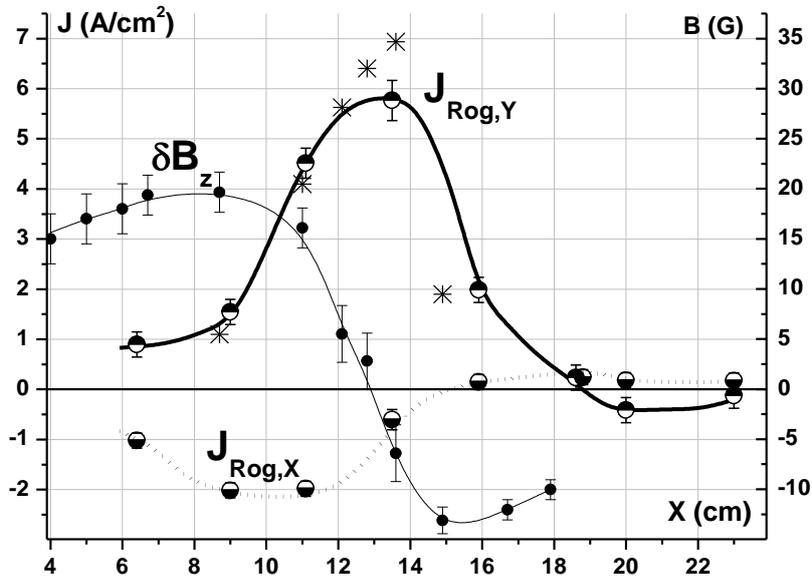

Figure 4. Profiles of magnetic field variation (●) measured by magnetic probe and two components of electric current density measured by Rogovski coil along X axis at Z,Y=0. By (✱) electric current density calculated from magnetic measurements is shown.

Measured magnetopause position $R_m$, a point where $\delta B_Z$ crosses zero, is at about 13.5 cm, which is significantly farther than the calculated stand off distance $R_M \approx 10\,\text{cm}$. The jump of field across boundary layer $\approx 50\,\text{G}$ is capable to stop the flow with velocity of only 55 km/s. In other words, magnetic pressure at magnetopause is about $2 \div 4$ times smaller than the kinetic pressure. Thus, ions should penetrate through magnetopause with only small deceleration, and the probe indeed measured ion flux deep inside of magnetosphere. All these features are very much similar to what have been observed by authors in previous experiments on mini-magnetosphere (*Shaikhislamov et al 2013, 2014*).

New experimental information comes from Rogovski coil measurements. In the fig. 4 two components of time-averaged current density are shown. The component corresponding to Chapman-Ferraro current well coincides with boundary layer and its value well agrees with approximate calculation from magnetic field data $J_y \approx (c/4\pi) \cdot dB_z/dx$. There is also a component of electric current along the plasma flow direction $J_x$. In previous experiments its value was deduced from non-coplanar magnetic field $J_x \approx -(c/4\pi)\partial B_y/\partial z$ and, due to Hall nature of $B_y$ component, was associated with the Hall current which cancels the advection electric field $E_y \approx (V_x - J_x/n_e e)B_z/c \approx 0$. The spatial profile obtained in the present work makes important verification that the Hall current is present only inside of mini-magnetosphere and its maximum value of $J_{electric} \approx 2.2$ A/cm$^2$ is rather close to the time-averaged upstream ion current $J_{ion} \approx 2.5$ A/cm$^2$.

Next come measurements at the flanks of magnetosphere. The probes have been positioned in terminator plane above and below equator plane $Z = \pm 3.5$ cm, and by movement along Y axis mapped magnetic structure. It is presented in fig. 5 for all three components of magnetic field variation. As expected, the $\partial B_y$ and $\partial B_x$ components have been found to be asymmetric in respect to equatorial plane, while $\partial B_z$ component was symmetric. Therefore, data only corresponding to location above equator $Z = 3.5$ cm are shown. At the east flank the structure is rather simple – compression of the main component $B_z$ inside of magnetosphere, boundary layer 5 cm in width and magnetopause location at about 11 cm. The variation of $\partial B_y$ component also reflects the compression of dipole field lines. In other words, the signs of $\partial B_z$ and $\partial B_y$ components correspond to the magnetospheric current directed tail-ward. A sun-ward component $B_x$ rises within boundary layer and corresponds in sign to differential tail-ward stretching of dipole field lines.

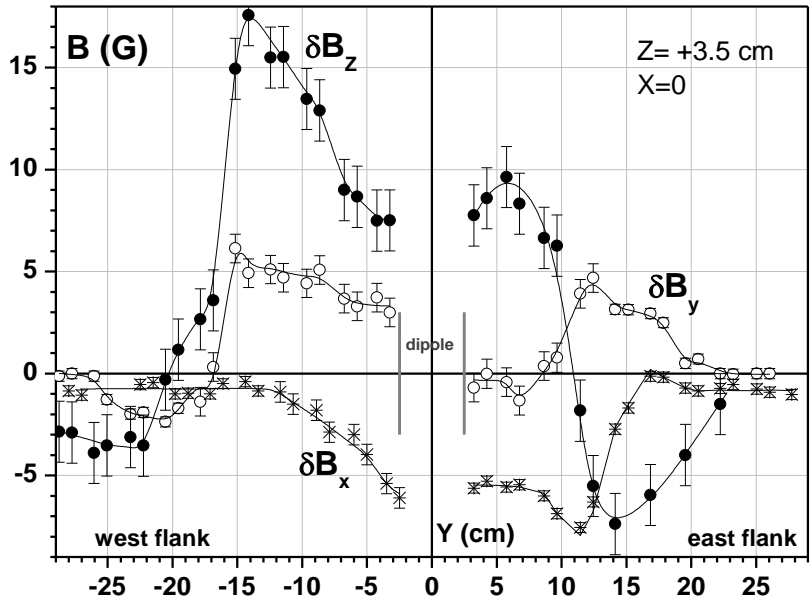

Figure 5. Profiles of magnetic field variations $\partial B_z$ (●), $\partial B_y$ (○) and $\partial B_x$ (✶) measured by magnetic probe along Y axis at Z=3.5 cm.

At the west flank the picture is only qualitatively similar. The quantitative differences consist of the following. Magnetopause is located comparatively much farther from the dipole at about 17-20 cm. Variation of the main component shows significant larger maximum value and correspondingly strong decrease inside of magnetosphere indicating the return current. There is significant $\partial B_y$ component inside of magnetosphere also corresponding in sign to the return current. Besides, no $\partial B_x$ is observed in the boundary layer as though dipole field lines are not sheared by differential tail-ward plasma motion.

The profile of plasma density in terms of measured ion current is shown in fig. 6. Once again, at the east flank the ion current drops within boundary layer and there is clear cavity devoid of plasma. Contrary to this, the whole west flank is filled by plasma and ion current collected by probe is even larger than the upstream value. Dynamic signals show that plasma moves though the west flank practically unimpeded and without deceleration. Presented data on magnetic field and plasma density verify in all details the results of previous experiments on asymmetry of mini-magnetosphere. The new and crucial information comes from direct measurements of current in plasma and its comparison with ion current. The X-component of electric current shows at the east flank a well defined current layer, which well agrees with computation of current density by magnetic field $J_x \approx (c/4\pi) \cdot (dB_z/dy - B_y/z)$, also shown in the fig. 6. At the west flank the current layer is wider and positioned much farther. Inside of the west flank there is a localized return current. Its maximum value is very close to the upstream ion current and to the value of ion current actually measured at this location. Thus, it can be deduced that the return current is carried by ions flowing past dipole. Note that the directly measured reversible current structure at the west flank is well supported by calculation using magnetic field data.

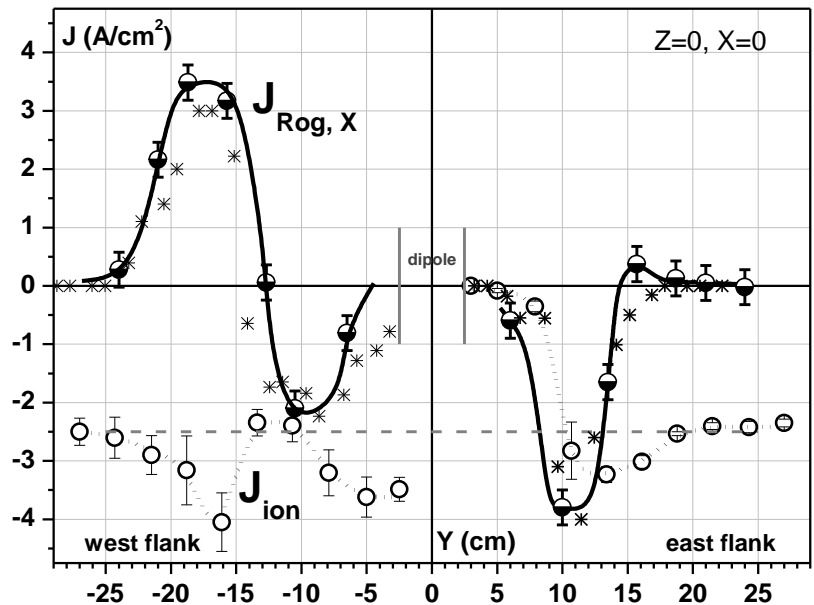

Figure 6. Profiles of X-component of electric current (●) measured by Rogovski coil and ion current (○) measured by Langmuir probe along Y axis. By (✻) X-component of electric current density calculated from magnetic measurements is shown. Gray dashed line indicates upstream value of ion current.

Finally, fig. 7 shows the profile of the Y-component of electric current. At the east flank there is current localized within the boundary layer which in sign, and within a factor of two in value, corresponds to deflected ions, as shown in fig. 1B for $V_y$ component of velocity. Obviously, such Hall current can exists only in the boundary layer where dipole field is present and should be zero outside of magnetosphere. At the west flank $J_y$ shows more complex reversible structure once again qualitatively reflecting the distribution of $V_y$ in fig. 1B. Thus, we can deduce that inside of magnetosphere and in the east boundary layer the gyrorotation of ions in dipole field produces electric current along Y axis.

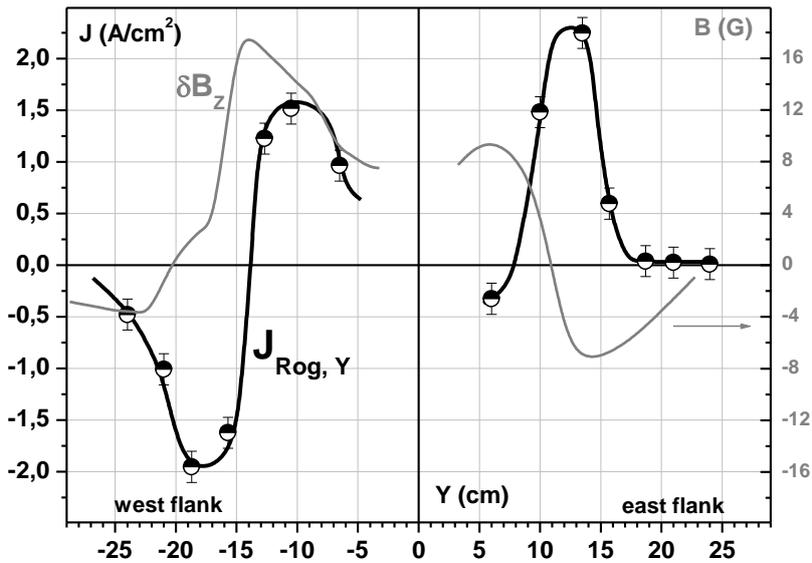

Figure 7. Profile of Y-component of electric current (●) measured by Rogovski coil along Y axis. Gray line indicates profile of $\partial B_z$ variation.

### 4. Conclusions

The aim of our study of the west-east asymmetry of mini-magnetosphere was to obtain experimental evidence of its Hall nature. Namely, that currents flowing inside of the mini-magnetosphere, and especially the return current at the west flank, are produced by ion gyrorotation in the dipole field. Because electrons are frozen into dipole field and are at rest, such ion current corresponds to zero induction electric field and to the Hall term being equal to the convection term, hence its Hall nature. First, by magnetic and plasma measurements we repeated the results of previous experiments on the main asymmetry features. At the east flank well defined boundary layer and plasma cavity were observed. Contrary to these, plasma was found to penetrate inside of the west flank and to support the return current, which was opposite in direction and comparable in value to the main Chapman-Ferraro current in the boundary layer. Direct measurements of electric current in plasma by Rogovski coil revealed that the return current is, in fact, equal in value to ion current, measured independently by Langmuir probe upstream of the dipole and inside of the west flank as well. Approximately of the same value electric current directed in the sense of ion gyrorotation have been observed at both flanks of mini-magnetosphere. Thus, the nature of the mini-magnetosphere asymmetry has been explained for the first time by Hall effects and illustrated by test particle model.

**Acknowledgements**
This work was supported by SB RAS Research Program grant II.8.1.4, Russian Fund for Basic Research grants 12-02-00367 and 14-29-06036, OFN RAS Research Program 15 and Presidium RAS Research Program 22.